\documentclass[aps,prl,twocolumn,showpacs,superscriptaddress,groupedaddress]{revtex4} 
\usepackage{graphicx}  % needed for figures
\usepackage{dcolumn}   % needed for some tables
\usepackage{bm}        % for math
\usepackage{amssymb}   % for math
\usepackage{amsfonts}
\usepackage{booktabs}
\usepackage{siunitx}
\usepackage{xcolor}
\usepackage{multirow}
\usepackage[normalem]{ulem}
\usepackage{amsmath}

\newcommand\tstrut{\rule{0pt}{3.0ex}}
\newcommand\bstrut{\rule[-1.5ex]{0pt}{0pt}}

\begin{document}

\title{Beyond Sphericity in a Semi-Magic Nucleus: Multiple-Shape Coexistence in $^{116}$Sn}
\author{M.~Siciliano}
    \email{msiciliano@anl.gov}
    \affiliation{Physics Division, Argonne National Laboratory, Lemont (IL), United States.}
	\affiliation{DPhN/Irfu/CEA, Universit\'e Paris-Saclay, Gif-sur-Yvette, France.}
	\affiliation{INFN, Laboratori Nazionali di Legnaro, Legnaro, Italy.}
\author{M.~Rocchini}
    \affiliation{INFN, Sezione di Firenze, Firenze, Italy.}
    \affiliation{Department of Physics, University of Guelph, Guelph, Canada.}
\author{T.R.~Rodr\'iguez}
    \email{trodrig@us.es}
    \affiliation{Departamento de F\'isica At\'omica, Molecular y Nuclear, Universidad de Sevilla, Sevilla, Spain.}
\author{A.~Goasduff}
	\affiliation{INFN, Laboratori Nazionali di Legnaro, Legnaro, Italy.}
	% \affiliation{Dipartimento di Fisica e Astronomia, Universit\`a di Padova, Padova, Italy.}
	% \affiliation{INFN, Sezione di Padova, Padova, Italy.}
\author{A.~Illana} %replied
	\affiliation{INFN, Laboratori Nazionali di Legnaro, Legnaro, Italy.}
    % \affiliation{Department of Physics, University of Jyv\"askyl\"a, Jyv\"askyl\"a, Finland.}
    \affiliation{Instituto de Estructura de la Materia, Consejo Superior de Investigaciones Cient\'ificas, Madrid, Spain.}
\author{M.~Saxena}
    \affiliation{University of Ohio, Athens (OH), United States.}
    \affiliation{Heavy Ion Laboratory, University of Warsaw, Warsaw, Poland.}
\author{G.~de~Angelis}
	\affiliation{INFN, Laboratori Nazionali di Legnaro, Legnaro, Italy.}
\author{S.~Bakes}
	\affiliation{INFN, Laboratori Nazionali di Legnaro, Legnaro, Italy.}
	\affiliation{Dipartimento di Fisica e Astronomia, Universit\`a di Padova, Padova, Italy.}
    \affiliation{Department of Physics, University of Surrey, Guildford, United Kingdom.}
\author{T.~Bayram} %replied
	\affiliation{INFN, Laboratori Nazionali di Legnaro, Legnaro, Italy.}
    \affiliation{Department of Physics, Karadeniz Technical University, Trabzon, T\"{u}rkiye.}
\author{D.~Bazzacco}
	\affiliation{INFN, Sezione di Padova, Padova, Italy.}
\author{K.~Belvedere}
    \affiliation{Department of Physics, University of Surrey, Guildford, United Kingdom.}
\author{G.~Benzoni}
    \affiliation{INFN, Sezione di Milano, Milano, Italy.}
\author{D.~Brugnara}
	\affiliation{INFN, Laboratori Nazionali di Legnaro, Legnaro, Italy.}
	\affiliation{Dipartimento di Fisica e Astronomia, Universit\`a di Padova, Padova, Italy.}
\author{P.~Colovi\'c}
    \affiliation{Ru{d\llap{\raise 1.22ex\hbox{\vrule height 0.09ex width 0.4em}}\rlap{\raise 1.22ex\hbox{\vrule height 0.09ex width 0.04em}}}er Bo\v{s}kovi\'{c} Institute and University of Zagreb, Zagreb, Croatia.}
\author{M.L.~Cortes}
	\affiliation{INFN, Laboratori Nazionali di Legnaro, Legnaro, Italy.}
%\author{I.~Djanto} %undeliverable
%    \affiliation{Department of Chemistry, Simon Fraser University, Burnaby (BC), Canada}
\author{D.T.~Doherty}
    \affiliation{Department of Physics, University of Surrey, Guildford, United Kingdom.}
\author{R.~Escudeiro}
	\affiliation{Dipartimento di Fisica e Astronomia, Universit\`a di Padova, Padova, Italy.}
	\affiliation{INFN, Sezione di Padova, Padova, Italy.}
\author{A.~Gadea}
    \affiliation{IFIC-CSIC, Universidad de Valencia, Valencia, Spain.}
\author{F.~Galtarossa} %replied
	%\affiliation{INFN, Laboratori Nazionali di Legnaro, Legnaro, Italy.}
    %\affiliation{IJCLab, Universit\'e Paris-Saclay, Orsay, France.}
	\affiliation{INFN, Sezione di Padova, Padova, Italy.}
\author{P.E.~Garrett}
    \affiliation{Department of Physics, University of Guelph, Guelph, Canada.}
    \affiliation{Department of Physics, University of the Western Cape, Bellville, South Africa.}
\author{N.~Gelli}
    \affiliation{INFN, Sezione di Firenze, Firenze, Italy.}
\author{E.T.~Gregor}
	\affiliation{INFN, Laboratori Nazionali di Legnaro, Legnaro, Italy.}
\author{A.~Gottardo}
	\affiliation{INFN, Laboratori Nazionali di Legnaro, Legnaro, Italy.}
\author{A.~Gozzelino} %replied
	\affiliation{INFN, Laboratori Nazionali di Legnaro, Legnaro, Italy.}
\author{J.~Ha} %replied
	\affiliation{Dipartimento di Fisica e Astronomia, Universit\`a di Padova, Padova, Italy.}
	\affiliation{INFN, Sezione di Padova, Padova, Italy.}
\author{K.~Hady\'{n}ska-Kl\c{e}k}
    \affiliation{Department of Physics, University of Surrey, Guildford, United Kingdom.}
    \affiliation{Heavy Ion Laboratory, University of Warsaw, Warsaw, Poland.}
%\author{S.Jazrawi} %undeliverable
%    \affiliation{Department of Physics, University of Surrey, Guildford, United Kingdom.}
\author{P.R.~John}
	\affiliation{Dipartimento di Fisica e Astronomia, Universit\`a di Padova, Padova, Italy.}
	\affiliation{INFN, Sezione di Padova, Padova, Italy.}
\author{C.E.~Jones}
    \affiliation{University of Brighton, Brighton, United Kingdom.}
\author{M.L.~Jurado-Gom\'ez}
    \affiliation{IFIC-CSIC, Universidad de Valencia, Valencia, Spain.}
%\author{H.M.~Jutila} %undeliverable
%    \affiliation{Department of Physics, University of Jyv\"askyl\"a, Jyv\"askyl\"a, Finland.}
\author{M.~Kici\'nska-Habior} %replied
    \affiliation{Department of Physics, University of Warsaw, Warsaw, Poland.}
\author{M.~Komorowska}
    \affiliation{Heavy Ion Laboratory, University of Warsaw, Warsaw, Poland.}
% \author{S.M.~Lenzi} %replied
% 	\affiliation{Dipartimento di Fisica e Astronomia, Universit\`a di Padova, Padova, Italy.}
% 	\affiliation{INFN, Sezione di Padova, Padova, Italy.}
\author{N.~Marchini}
    \affiliation{Dipartimento di Fisica, Universit\'a di Camerino, Camerino, Italy.}
    \affiliation{INFN, Sezione di Firenze, Firenze, Italy.}
\author{M.~Matejska-Minda} %replied
	\affiliation{Institute of Nuclear Physics Polish Academy of Sciences, Krakow, Poland.}
\author{R.~Menegazzo}
	\affiliation{INFN, Sezione di Padova, Padova, Italy.}
\author{D.~Mengoni}
	\affiliation{Dipartimento di Fisica e Astronomia, Universit\`a di Padova, Padova, Italy.}
	\affiliation{INFN, Sezione di Padova, Padova, Italy.}
% \author{A.~Montaner-Piz\'a} %no mail
% 	\affiliation{Dipartimento di Fisica e Astronomia, Universit\`a di Padova, Padova, Italy.}
% 	\affiliation{INFN, Sezione di Padova, Padova, Italy.}
\author{A.~Nannini}
    \affiliation{INFN, Sezione di Firenze, Firenze, Italy.}
\author{P.J.~Napiorkowski}
    \affiliation{Heavy Ion Laboratory, University of Warsaw, Warsaw, Poland.}
\author{D.R.~Napoli}
	\affiliation{INFN, Laboratori Nazionali di Legnaro, Legnaro, Italy.}
\author{J.~Ojala} %replied
    \affiliation{Department of Physics, University of Jyv\"askyl\"a, Jyv\"askyl\"a, Finland.}
\author{G.~Pasqualato}
	\affiliation{Dipartimento di Fisica e Astronomia, Universit\`a di Padova, Padova, Italy.}
	\affiliation{INFN, Sezione di Padova, Padova, Italy.}
\author{C.M.~Petrache} %replied
    \affiliation{IJCLab, Universit\'e Paris-Saclay, Orsay, France.}
    \affiliation{Department of Chemistry, Simon Fraser University, Burnaby (BC), Canada.}
    \affiliation{Institute of Modern Physics, Chinese Academy of Science, Lanzhou, China.}
\author{F.~Recchia}
	\affiliation{Dipartimento di Fisica e Astronomia, Universit\`a di Padova, Padova, Italy.}
	\affiliation{INFN, Sezione di Padova, Padova, Italy.}
\author{D.~Testov} %replied
    \affiliation{Dipartimento di Fisica e Astronomia, Universit\`a di Padova, Padova, Italy.}
	\affiliation{INFN, Sezione di Padova, Padova, Italy.}
   	\affiliation{Extreme Light Infrastructure-Nuclear Physics, Horia Hulubei National Institute for Physics and Nuclear Engineering, Magurele, Romania.}
\author{J.J.~Valiente-Dob\'on}
	\affiliation{INFN, Laboratori Nazionali di Legnaro, Legnaro, Italy.}
    \affiliation{IFIC-CSIC, Universidad de Valencia, Valencia, Spain.}
\author{I.~Zanon}
	\affiliation{INFN, Laboratori Nazionali di Legnaro, Legnaro, Italy.}
    \affiliation{Dipartimento di Fisica e Scienze della Terra, Universit\'a di Ferrara, Ferrara, Italy.}
% \author{G.~Zhang} %replied
%     \affiliation{Dipartimento di Fisica e Astronomia, Universit\`a di Padova, Padova, Italy.}
% 	\affiliation{INFN, Sezione di Padova, Padova, Italy.}
\author{M.~Zieli\'nska}
	\affiliation{DPhN/Irfu/CEA, Universit\'e Paris-Saclay, Gif-sur-Yvette, France.}

\begin{abstract}
The study of nuclear shape evolution and coexistence provides key insight into the nuclear interaction, particularly in semi-magic systems expected to be spherical. 
A high-precision Coulomb-excitation measurement of $^{116}$Sn yields a comprehensive set of electromagnetic matrix elements, enabling the determination of quadrupole moments of the $2_{1,2,3}^+$ states and intrinsic deformations of the $0_{1,2,3}^+$ states. 
The results provide an unambiguous and direct proof of the rare phenomenon of multiple-shape coexistence, revealing a weakly deformed ground state incompatible with the spherical shape.
\end{abstract}

\maketitle

Experimental evidence for multiple-shape coexistence has been obtained in only a handful of nuclei across the nuclear chart~\cite{andreyev2000, leoni2017ni, PhysRevLett.121.192501, CRUZ201894, garrett2019multiple, siciliano2020coexistence, ojala2022pb}.
In these cases, the occurrence of three or more low-lying $0^+$ states with distinct intrinsic deformations and mutual mixing challenges both experimental sensitivity and theoretical modeling. 
Nuclei near closed shells offer an exceptional testing ground for such phenomena, as they allow the interplay between spherical shell-model configurations and emergent collectivity to be explored in a controlled manner. 
Semi-magic systems, long regarded as structurally simple due to their proximity to spherical shell closures, therefore provide ideal laboratories for probing the limits of single-particle and collective dynamics and for constraining the nuclear interaction~\cite{Pauling1965magic, siciliano2020plb, Mougeot100Sn, Henderson208Pb}.

Within this framework, the Sn isotopes ($Z\!=\!50$) have traditionally been viewed as paradigmatic examples of seniority-dominated structure, where low-lying excitations arise from nucleon-pair breaking within a largely inert core~\cite{ressler2004transition, talmi1971generalized}. 
This picture, supported by strong pairing correlations, accounts for several key systematics, including the nearly constant excitation energies of the $2_1^+$ and $4_1^+$ states and the presence of low-lying isomers over a wide range of neutron numbers~\cite{astier2012seniority, maheshwari2016asymmetric, kumar2017seniority, maheshwari2019seniority}. 
However, an increasing body of experimental evidences highlight important limitations of purely seniority-based interpretations. 
The existence of low-lying excited $0^+$ states connected by strong electric monopole ($E0$) transitions~\cite{BACKLIN1981490, cross2017}, together with their selective population in nucleon-transfer reactions~\cite{YAGI1968129, FLEMING19701, FIELDING1977389, BRON1979335, PhysRevC.85.054609, shapecoexistence2022}, points to a more complex structure involving multi-particle–multi-hole intruder configurations.
Historically, this complexity has been interpreted in terms of a spherical ``normal'' configuration coexisting and mixing with a single deformed intruder band, commonly associated with proton excitations across the $Z\!=\!50$ shell gap~\cite{pore2017, cross2017, spieker2018shape, petrache2019}. 
Yet the characteristic V-shaped systematics of the excitation energies already suggest that this description is incomplete, since both $0_{2,3,4}^+$~\cite{LEONI2024104119} and higher-spin states~\cite{shapecoexistence2022} show such a characteristic intruding behavior. 
This picture is reinforced by the strong population of excited $0^+$ states in ($^3$He,n) reactions~\cite{FIELDING1977389}, indicating distinct configurations with significant $(2p-2h)_\pi$ character, which was then supported by the spectroscopic study of Refs.~\cite{pore2017, petrache2019}. 
The ($^4$He,2n$\gamma$) spectroscopy of Ref.~\cite{BRON1979335}, instead, identified a rotational-like band built on the $0_2^+$ state and associated it with a deformation driven by the opening of a proton Nilsson gap. 
Together with the observation of enhanced $E0$ strengths among low-lying $0^+$ states~\cite{BACKLIN1981490, cross2017} suggesting strong configuration mixing, these findings provide compelling evidence that the low-energy structure of Sn isotopes is governed by multiple-shape coexistence, rather than a single normal and intruder configuration. 

A model-independent identification of such coexisting structures requires highly selective probes. 
Coulomb excitation is uniquely suited for this purpose, as it enables precise determinations of $E2$ matrix elements and thus direct access to intrinsic shapes and collectivity~\cite{zielinska2022euroschool}. 
Directly observing multiple-shape coexistence in a semi-magic nucleus is particularly significant, as it challenges the conventional distinction between ``simple'' and ``collective'' systems and calls for a reassessment of nuclear structure in regions long thought to be well understood. 
Despite its importance, extensive multi-step Coulomb-excitation data in the Sn mass region have remained scarce, especially at the level required to disentangle multiple coexisting configurations. 

In this Letter, we present the results of a comprehensive Coulomb-excitation experiment on $^{116}$Sn designed to probe shape coexistence in a model-independent manner. 
By combining the present data with existing spectroscopic information and state-of-the-art beyond-mean-field calculations within the Projected Generator Coordinate Method, we provide a comprehensive reinterpretation of the low-energy structure of $^{116}$Sn.
The comparison with theory highlights both the predictive power and the present limitations of microscopic approaches in describing shape coexistence in semi-magic nuclei.  
Our findings firmly establish multiple shape coexistence in a semi-magic nucleus and underscore the importance of this phenomenon for understanding the nuclear interaction.

\begin{figure}[t!]
    \centering    
    \includegraphics[width=0.49\textwidth]{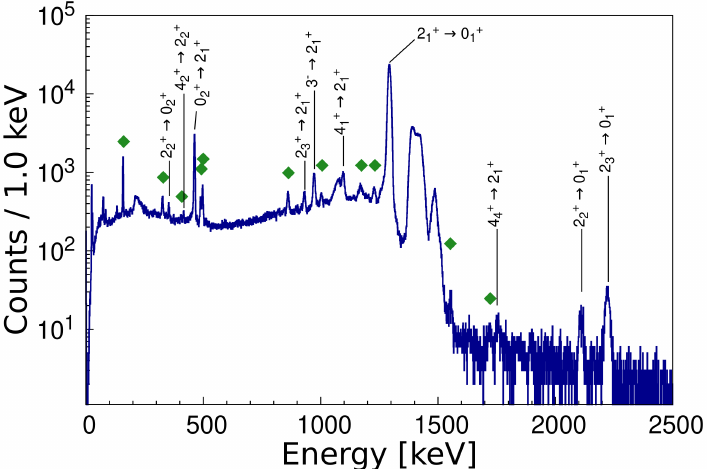}
    \vspace{-7mm}
    \caption{\label{fig:Spectrum} (color online) Compton-suppressed $\gamma$-ray energy spectrum for the $^{58}$Ni+$^{116}$Sn Coulomb-excitation reaction obtained by requiring the identification of the ejectiles in SPIDER. The $\gamma$-ray energy is Doppler corrected for the $^{116}$Sn recoil kinematics. The identified $\gamma$-ray transitions of $^{116}$Sn are labeled, while the green diamonds indicate the Sn-isotope contaminants in the target material.}
\end{figure}

The low-lying structure of $^{116}$Sn was investigated at the INFN–Legnaro National Laboratories through a multi-step Coulomb-excitation experiment. 
A continuous 190-MeV $^{58}$Ni beam from the XTU-Tandem accelerator impinged on a self-supporting 2.3-mg/cm$^2$ $^{116}$Sn target enriched to 98.0(1)\%. 
The beam energy was chosen to maximize excitation probability while fulfilling Cline’s safe-distance criterion~\cite{gosia1986}, ensuring a purely electromagnetic interaction. 
De-excitation $\gamma$ rays were detected with the GALILEO spectrometer~\cite{galileo}, comprising 20 Compton-suppressed HPGe detectors, while backscattered beam ions were measured by the SPIDER silicon detector array~\cite{spider}. 
This geometry enhances multi-step excitation and sensitivity to quadrupole moments and electromagnetic matrix-element signs, making it optimal for probing nuclear shapes. 
The measured energies and angles of the scattered $^{58}$Ni ions allowed reconstruction of the $^{116}$Sn recoil kinematics and accurate Doppler correction of the $\gamma$-ray energies.
Figure~\ref{fig:Spectrum} shows the $\gamma$-ray spectrum obtained by summing all GALILEO detectors in coincidence with SPIDER. 
Details of the data presorting and analysis procedures are reported in Refs.~\cite{AR_116Sn, rocchini66zn, marchini94zr}.

\begin{figure}[t!]
     \centering
     \includegraphics[width=0.48\textwidth]{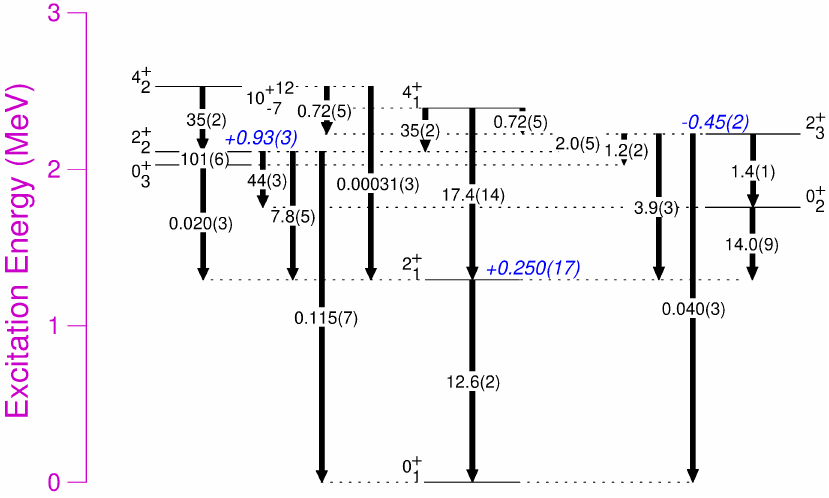}
     \vspace{-7mm}
     \caption{\label{fig:ExpLevelScheme} (color online) Partial level scheme of $^{116}$Sn, showing the three coexisting structures discussed in this Letter. Arrow labels indicate the $B(E2)$ transition strength in Weisskopf units, while the spectroscopic quadrupole moments of the $2_{1,2,3}^+$ states, expressed in $e$b, are shown in blue and italic.}
\end{figure}

Electromagnetic matrix elements (MEs) were extracted from the Coulomb-excitation data by analyzing the evolution of $\gamma$-ray yields as a function of particle scattering angle. 
The yields were normalized to the $2_1^+ \to 0_1^+$ transition, using the well-known lifetime $\tau(2_1^+) = 0.540(14)$~ps in $^{116}$Sn~\cite{NuclData116Sn}. 
Since excited states may be populated through multiple excitation paths, the measured cross sections depend nonlinearly on several MEs. 
To enhance sensitivity to angular distributions and disentangle individual contributions, the data were divided into nine sub-datasets corresponding to the eight SPIDER rings and the full array. 
The resulting set of $E2$ MEs is reported in Table~\ref{tab:ME}, while the corresponding reduced transition probabilities $B(E2)$ and spectroscopic quadrupole moments $Q_s$ for the low-lying $2^+$ states are shown in Fig.~\ref{fig:ExpLevelScheme}.
These results enable a direct characterization of the intrinsic shapes of the coexisting $0_{1,2,3}^+$ states, providing decisive evidence for multiple-shape coexistence in the $Z=50$ region. 
In particular, the large positive value of $Q_s(2_1^+)$ immediately points to an oblate ground-state deformation. 
Such a result is remarkable since only $^{116}$Sn and $^{110}$Sn~\cite{park110Sn} exhibit sizable positive quadrupole moments, whereas the remaining Sn isotopes are characterized by $Q_s(2_1^+)\!\approx\!0$, consistent with the traditional picture of nearly spherical semi-magic nuclei~\cite{allmond2015investigation}.

\begin{table}[t!]
\caption{\label{tab:ME} $E2$ matrix elements extracted in the present analysis, presented together with the corresponding $B(E2; J_i^\pi \to J_f^\pi)$ transition strengths and spectroscopic quadrupole moments $Q_s$. The relative signs of the matrix elements are reported with respect to the matrix elements whose sign is listed in brackets, assumed to be positive. The reported uncertainties have been increased by additional 3\% to account for systematic effects inherited by GOSIA analysis~\cite{zielinska2016gosia}.}
\centering
\begin{tabular}{c|c|c}
\hline\hline
$J_i^\pi \to J_f^\pi$   &   $\langle J_i^\pi | \hat{E2} | J_f^\pi \rangle$ [e\,b]   &   $B(E2)$ [W.u.]  \tstrut\bstrut \\
\hline
$2_1^+ \to 0_1^+$       &   $(+)0.460~(3)$                  &   $12.6~(2)$         \tstrut \\
$0_2^+ \to 2_1^+$       &   $+0.217~(7)$                    &   $14.0~(9)$         \\
$0_3^+ \to 2_1^+$       &   $-0.00813~(56)$                 &   $0.020~(3)$        \\
$2_2^+ \to 0_3^+$       &   $-1.30~(4)$                     &   $101~(6)$          \\
$2_2^+ \to 0_2^+$       &   $-0.86~(3)$                     &   $44~(3)$           \\
$2_2^+ \to 2_1^+$       &   $(+)0.363~(12)$                 &   $7.8~(5)$          \\
$2_2^+ \to 0_1^+$       &   $-0.044~(1)$                    &   $0.115~(7)$        \\
$2_3^+ \to 2_2^+$       &   $-0.181_{-0.020}^{+0.026}$      &   $2.0~(5)$          \tstrut\bstrut \\
$2_3^+ \to 0_3^+$       &   $+0.144~(11)$                   &   $1.2~(2)$          \\
$2_3^+ \to 0_2^+$       &   $+0.153~(7)$                    &   $1.4~(1)$          \\
$2_3^+ \to 2_1^+$       &   $(+)0.256~(9)$                  &   $3.9~(3)$          \\
$2_3^+ \to 0_1^+$       &   $-0.0260~(9)$                   &   $0.040~(3)$        \\
$4_1^+ \to 2_3^+$       &   $+1.01_{-0.08}^{+0.05}$         &   $34_{-5}^{+4}$     \tstrut\bstrut \\
$4_1^+ \to 2_2^+$       &   $+1.019~(37)$                   &   $34~(2)$           \\
$4_1^+ \to 2_1^+$       &   $(+)0.726~(28)$                 &   $17.4~(14)$        \\
$4_2^+ \to 4_1^+$       &   $(+)0.55_{-0.24}^{+0.27}$       &   $10_{-7}^{+12}$    \tstrut\bstrut \\
$4_2^+ \to 2_3^+$       &   $+0.147~(5)$                    &   $0.72~(5)$         \\
$4_2^+ \to 2_2^+$       &   $+1.030~(33)$                   &   $35~(2)$           \\
$4_2^+ \to 2_1^+$       &   $+0.00307~(14)$                 &   $0.00031~(3)$      \\
\hline\hline
$J^\pi$   &   $\langle J^\pi | \hat{E2} | J^\pi \rangle$ [e\,b]   &   $Q_s$ [e\,b]   \tstrut\bstrut \\
\hline
$2_1^+$       &   $+0.33~(2)$        &   $+0.250~(17)$  \tstrut \\
$2_2^+$       &   $+1.2~(4)$         &   $+0.93~(3)$            \\
$2_3^+$       &   $-0.60~(3)$        &   $-0.45~(2)$    \bstrut \\
\hline\hline
\end{tabular}
\end{table}

\begin{figure}[t!]
    \centering
    \includegraphics[width=0.45\textwidth]{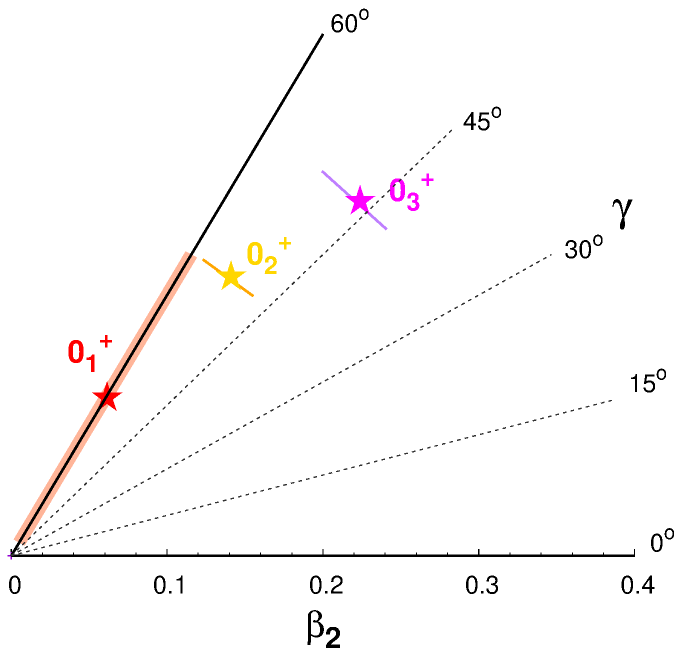}
    \vspace{-5mm}
    \caption{\label{fig:deformations} (color online) Quadrupole-deformation parameters of the $0_1^+$ (red star), $0_2^+$ (yellow star), and $0_3^+$ (purple star) states in $^{116}$Sn. Color-matched lines indicate the $1\sigma$ uncertainties, while the orange rectangle denotes the $\delta\beta_2$ softness of the $0_1^+$ state.}
\end{figure}

Because the electric quadrupole operator is a spherical tensor, its zero-coupled products form rotational invariants that can be related to intrinsic deformation parameters via the quadrupole sum rules~\cite{kumar1972, gosia1986}. 
From the measured MEs, the lowest-order invariant $\langle Q^2 \rangle$, constructed from products of $E2$ matrix elements connecting a given state to all accessible final states, provides the overall quadrupole deformation $\beta_2$ as 
$
    \langle Q^2 \rangle\!=\!\frac{3}{4 \pi} Z R^2 \langle \beta_2 \rangle^2 \, ,
$
where $R\!=\!1.2\,A^{1/3}~fm$. 
The next-order invariant $\langle Q^3 cos(3\delta) \rangle$, obtained from triple products of $E2$ MEs, provides information on the parameter $\delta$ which is the departure from axial symmetry for the charge-state distribution. 
Such a rotational invariant is sensitive to the relative sign of the MEs and, under the assumption $\delta \!=\! \gamma$, describes the triaxiality $\gamma$ of a nucleus, yielding $\langle Q^3 cos(3\delta) \rangle \!\approx\! \langle Q^2 \rangle^{3/2} \langle cos(3\gamma) \rangle$.
Application of these invariants yields direct, model-independent information on the intrinsic shapes of the low-lying $0^+$ states. 
As shown in Fig.~\ref{fig:deformations}, the ground state of $^{116}$Sn is oblate, with $\beta_2=0.121(3)$. 
With increasing excitation energy, the deformation strength grows significantly: the $0_2^+$ and $0_3^+$ states exhibit $\beta_2=0.232(6)$ and $0.324(11)$, respectively, accompanied by increasing triaxiality, evolving from $\gamma=53(5)^\circ$ to $47(5)^\circ$.

Higher-order invariants probe fluctuations about the equilibrium shape. 
The fourth-order invariant, proportional to $\langle Q^4\rangle$, allows the variance of the deformation $\langle \delta\beta_2 \rangle$ to be evaluated. 
Owing to the limited population of higher-lying states, this invariant could be reliably extracted only for the ground state, yielding $\langle \delta\beta_2 \rangle=0.110(5)$. 
The extracted soft-oblate deformation of the ground state agrees with the predictions of Ref.~\cite{togashi2018sn}. 
This finding directly challenges the long-standing assumption that semi-magic Sn isotopes possess spherical ground states enforced by the closed $Z\!=\!50$ shell, since in $^{116}$Sn it actually exhibits a small but finite deformation with limited shape fluctuations.

%%%%%%%%%%%%%%%%%%%%%%%%%%%%%%%%%%%%%%%%%%%%%%%%%%%%%%%%%
%%%%%%%%%%% PGCM %%%%%%%%%%%%%%%%%%%%%%%%%%%%%%%%%%%%%%%%
%%%%%%%%%%%%%%%%%%%%%%%%%%%%%%%%%%%%%%%%%%%%%%%%%%%%%%%%%

The low-lying spectrum of $^{116}$Sn was calculated using the Gogny-D1S energy density functional~\cite{berger1984microscopic} within the Projected Generator Coordinate Method (PGCM) (see Ref.~\cite{Robledo2019} and references therein). 
This method has proven highly successful in describing shape evolution, mixing, and coexistence. 
In this framework, the nuclear wave functions are constructed as linear combinations of symmetry-restored Hartree–Fock–Bogoliubov (HFB) states characterized by different collective properties. 
Two complementary implementations of the method have been carried out: a standard one, in which static shapes are explored in the triaxial quadrupole deformation plane $(\beta_{2},\gamma)$. 
However, owing to the variational nature of the method, the ground-state energy is favored relative to excited states, and the resulting excitation spectrum is therefore expected to be somewhat stretched compared to the exact solution~\cite{Borrajo2015cranking, Egido2016, Rodriguez2020, Rodriguez2016cranking}. 
The second implementation expands the variational space for states with angular momentum $J\!\neq\!0$. 
In this PGCM calculation, axial --or quasi-axial-- HFB wave functions are introduced, featuring oblate-prolate quadrupole deformations characterized by $(\pm\beta_{2})$, which can intrinsically rotate around the $x-$axis with a cranking angular momentum, $J_{c}$. 
As a first step, intrinsic HFB states were generated using the particle-number–variation-after-projection (PNVAP) approach~\cite{anguiano2001particle} in the triaxial $(\beta_2,\gamma)$ plane. 
The resulting total-energy surface (not shown) exhibits two well-separated minima. 
The absolute minimum occurs at the spherical point $(\beta_2\!=\!0)$, corresponding to the semi-magic $Z\!=\!50$ configuration, while a secondary prolate minimum appears at $(\beta_2,\gamma)\!\approx\!(0.3,0^\circ)$ which originates from the opening of a proton Nilsson gap at finite deformation, driven by the upward-sloping orbitals stemming from the spherical $0g_{9/2}$ and $1f_{7/2}$ shells and the downward-sloping levels from the remaining $0g1d2s$ shell and the $0h_{11/2}$ orbital. 
This surface already suggests the coexistence of spherical and deformed configurations in $^{116}$Sn, although a direct comparison with experiment requires the full PGCM treatment.

\begin{figure}[b!]
     \centering
     \includegraphics[width=0.48\textwidth]{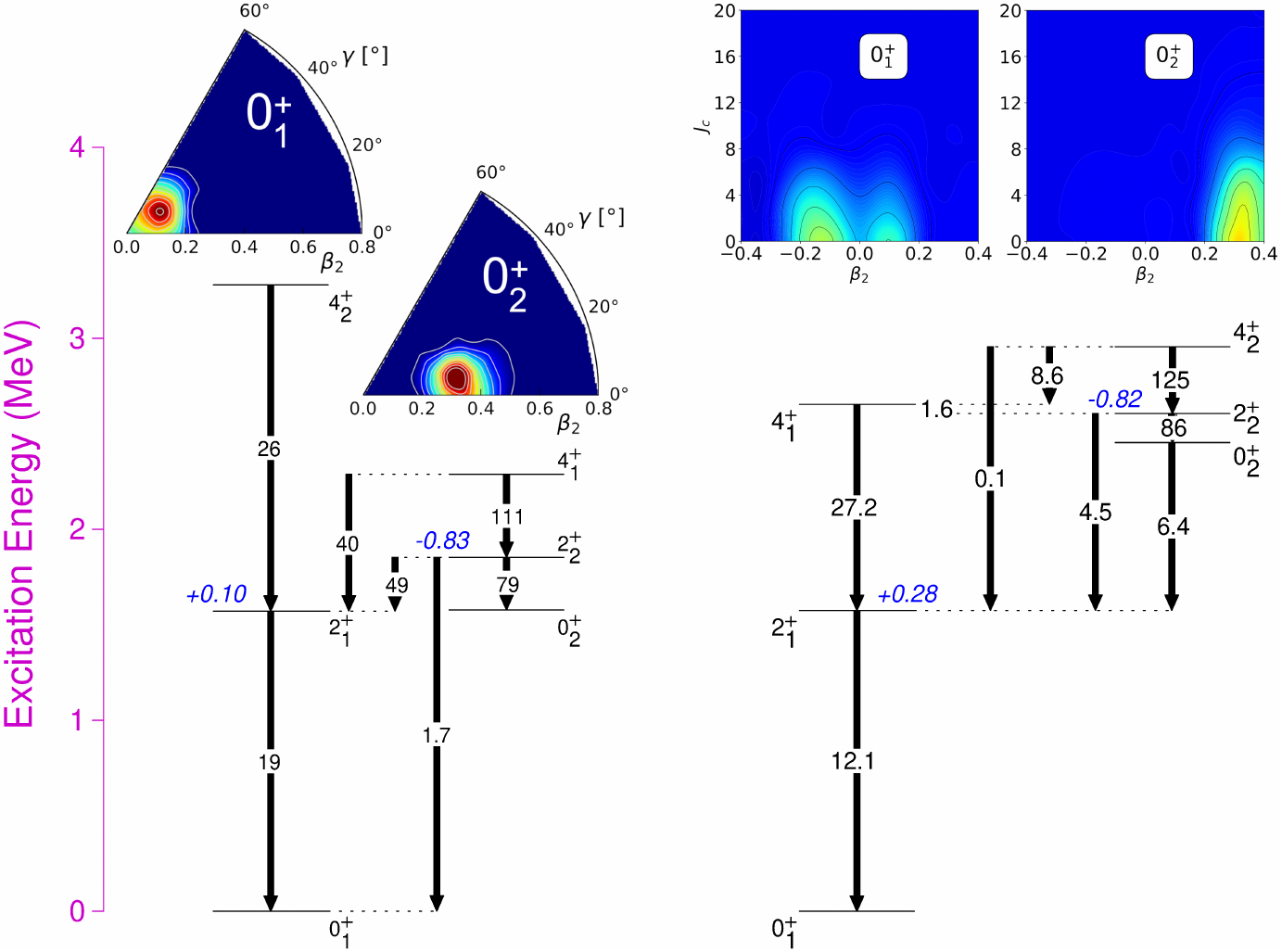}
     \vspace{-7mm}
     \caption{\label{fig:ThLevelScheme} (color online) Energy spectrum predicted by the PGCM calculation employing the Gogny-D1S energy density functional for the (left) triaxial quadrupole-deformation and (right) cranking angular-momentum approaches. Arrow labels indicate the $B(E2)$ transition strength in Weisskopf units, while the spectroscopic quadrupole moment of the $2_{1,2,3}^+$ states, expressed in $e\,b$, are shown in blue. The collective wave functions of the calculated $0_{1,2}^+$ states are displayed.}
\end{figure}

The intrinsic HFB states were therefore projected onto good particle number and angular momentum and subsequently mixed, resulting in the excitation spectrum and collective wave functions shown in Fig.~\ref{fig:ThLevelScheme}. 
Analyzing first the triaxial calculation, a ground-state band with nearly equidistant levels ($\approx\!1.5$~MeV) is predicted. 
The collective wave function of the ground state shows the largest contribution at small deformation and its maximum is located at $(\beta_{2},\gamma)=(0.15,30^{\circ})$. 
This indicates that this semi-magic nucleus does not reach its maximum contribution at the spherical point, which would otherwise correspond (except for pairing correlations) to the closed-shell configuration. 
Furthermore, the remaining states in the band exhibit an oblate structure. 
Hence, the in-band $B(E2)$ and the $Q_{s}(2^{+}_{1})$ values are in fair agreement with the experimental results.
A second collective band, built on the theoretical $0_2^+$ state, is associated with a relatively stable deformation, peaking near $(\beta_2,\gamma)\!\approx\!(0.3,10^\circ)$, close to the secondary minimum of the energy surface. 
Its experimental counterpart, however, cannot be uniquely identified: while the predicted large in-band $B(E2)$ strengths resemble those of the experimental band built on the $0_3^+$ state, the sizable negative $Q_s(2_2^+)$ instead may suggest the presence of a deformed structure built by the experimental $0_2^+$ and $2_3^+$ states. 
Nevertheless, the PGCM calculation predicts a prolate band, in contrast to the nearly oblate deformation inferred experimentally for both excited $0^+$ structures. 
This discrepancy suggests that important correlations remain absent from the present theoretical framework and illustrates the challenge of describing shape coexistence and configuration mixing in semi-magic nuclei within current beyond-mean-field approaches.
Beyond these two structures, the PGCM calculations also predicts a third band built on the $0_4^+$ state, which exhibits a larger deformation and increased triaxiality, with a maximum around $(\beta_2,\gamma)\!\approx\!(0.4,20^\circ)$. 
The calculations predict a nearly degenerate $0_3^+$ state close in energy to the $0_4^+$ state: its collective wave function is dominated by oblate–triaxial configurations and exhibits significant shape mixing with the triaxial region associated with the $0_4^+$ state. 
Although the ordering of the excited $0^+$ states differs from experiment, the calculation naturally generates three competing collective structures at low excitation energy, supporting the interpretation of $^{116}$Sn in terms of multiple shape coexistence. 

One of the primary shortcomings of the triaxial calculation is the significant stretching observed in the excitation energies of the ground-state band. 
The axial calculation, incorporating cranking wave functions, dramatically corrects this effect, bringing the energies of the $2^{+}_{1}$ and $4^{+}_{1}$ states much closer to experimental values (see Fig.~\ref{fig:ThLevelScheme}). 
Nevertheless, the interpretation of the nuclear structure remains similar based on the wave functions. 
The ground state is a mixture of $J_{c}=0$ oblate and prolate states, nearly symmetric around the spherical point, which is consistent with the projection of the triaxial collective wave function onto the $\gamma=0^{\circ}$ and $\gamma=60^{\circ}$ axes. 
The $2^{+}_{1}$, $4^{+}_{1}$, and higher-spin states are also oblate, with increasing $J_{c}$ components as the angular momentum rises. 
These components, absent in the triaxial calculation, are responsible for the compression of the spectrum. 
Finally, a prolate rotational band built upon the $0^{+}_{2}$ state also emerges. 

The results of this work reinforce the paradoxical nature of the Sn isotopic chain. 
For decades, these nuclei have been regarded as the archetypal spherical, pairing-dominated systems and the textbook case underpinning the seniority scheme. 
Yet, the present Coulomb-excitation study demonstrates that even the ground state of a semi-magic Sn isotope departs from spherical symmetry. 
More strikingly, the coexistence of multiple $0^+$ configurations with distinct intrinsic shapes, firmly established through the present measurement, directly proves that $^{116}$Sn exhibits multiple-shape coexistence --a phenomenon so rare that it has been identified in only a handful of nuclei across the nuclear chart. 
Microscopic calculations using the Gogny interaction show, in agreement with experimental data, a non-spherical ground state and several low-energy $0^{+}$ states with distinct shapes. 
However, they also predict a prolate rotational band that is at odds with the experimental rotational band, which exhibits an oblate character. 
This discrepancy emphasizes that the experimental observation of multiple shape coexistence in $^{116}$Sn provides a particularly stringent benchmark for microscopic theories, highlighting the need for an improved treatment of correlations in semi-magic nuclei. 
More broadly, the simultaneous emergence of ground-state deformation and multiple shape coexistence in a nucleus long regarded as a paradigm of spherical structure demonstrates that shell closures do not preclude the development of complex collective dynamics. 
Rather, even near closed shells, subtle correlations can profoundly reshape the nuclear landscape, making semi-magic nuclei a uniquely sensitive laboratory for constraining the nuclear interaction.

\section{Acknowledgment}
%The authors would like to thank the GALILEO collaborations. 
Special thanks go to the INFN-LNL technical staff for their help in setting up the apparatuses and the good quality beam. 
The authors acknowledge the contribution of the GAMMAPOOL European Gamma-Ray Spectroscopy Pool. 
The authors are also grateful to CloudVeneto~\cite{CloudVeneto} for the use of computing and storage facilities. 
This manuscript owes much to the collaboration with D.~Kalaydjieva. 
This research was partially supported by the European Union’s Seventh Framework Programme for Research and Technological Development (grant no. 262010). 
The author (MS) is supported by the OASIS project no. ANR-17-CE31-0026, 
and by the U.S. Department of Energy, Office of Science, Office of Nuclear Physics, under contract number DE-AC02-06CH11357. 
The author (TRR) acknowledges the Grants PID2021-127890NB-I00 and PID2024-159559NB-C22 funded by MCIN/AEI/10.13039/501100011033, and the support of the GSI-Darmstadt computer facilities. 
The authors (TB) acknowledge financial support from the Scientific and Technological Research Council of T\"{u}rkiye [T\"{U}B\.{I}TAK] under Grant No. 1059B191601912, 
(AG) from MICIU/AEI/10.13039/501100011033 Spain with grants PID2023-150056NB-C4, PID2024-156674NB-I00, CEX2023-001292-S, from Generalitat Valenciana Spain with grant CIPROM/2022/54, and from the EU FEDER funds, 
(MMM) from the National Science Centre of Poland under grant No.~2023/50/E/ST2/00621,
(JO) from the Research Council of Finland Grant No. 307685.

\bibliography{116Sn_Ref}

\newpage
\begin{figure*}[b!]
\includegraphics[width=\textwidth]{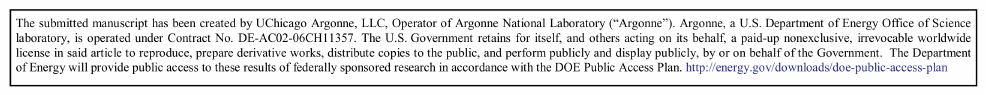}    
\end{figure*}

\end{document}